\documentclass[12pt]{article}
\pdfoutput=1
\usepackage{jheppub}
\usepackage{amsmath,amssymb}
\usepackage{graphicx}
\makeatletter

\@addtoreset{equation}{section}
\makeatother

\newcommand{\beq}{\begin{equation}}
\newcommand{\eeq}{\end{equation}}

\begin{document}

\baselineskip=18pt  % a la harvmac
\baselineskip 0.7cm

\begin{titlepage}

%% Set the number of the title with 0
\setcounter{page}{0}

% change the footnote symbol
\renewcommand{\thefootnote}{\fnsymbol{footnote}}

%------------------
\begin{flushright}
%preprint number
%CALT-mm-nnnn\\
%IPMU09-nnnn\\
%UT-09-30\\
%month, 2008
\end{flushright}
%---------------------

\vskip 1.5cm

\begin{center}
{\LARGE \bf
 $tt^*$ Geometry and a Twistorial Extension of Topological Strings
\vskip 1.5cm 
}

{\large
 Cumrun Vafa
\\
\medskip
}

\vskip 0.5cm

{
\small\it
Jefferson Physical Laboratory, Harvard University, Cambridge, MA 02138, USA
\\
\medskip
}
\end{center}

%-----------------------------------------
\centerline{{\bf Abstract}}
\medskip
\noindent

We extend the recent study of 3d $tt^*$ geometry to 5d (4d) ${\cal N}=1$ (${\cal N}=2)$ supersymmetric theories.
We show how the amplitudes of $tt^*$ geometry lead to an extension of topological
strings and Nekrasov partition functions to twistor space, where the north/south poles of the twistor sphere lead to
topological/anti-topological string amplitudes.   However, the equator of the twistor sphere is the physical region for the amplitudes, where the unitarity of physical amplitudes is respected.  We propose that the amplitudes for line operators studied by Gaiotto, Moore and Neitzke lead to
difference equations for the twistorial topological string partition function, extending the familiar one for the monodromy of Branes in the context of the standard topological strings.   This setup unifies
$tt^*$ geometry, topological strings and the hyperK\"ahler geometry formulated by GMN into a single framework.

\end{titlepage}
\setcounter{page}{1} % don't number title page

%%%%%%%%%%%%%%%%%%%%%%%%%%%%%%%%%%%%%%%%%%%%%%%%%%%%%%%%%%%%%%%%%%%%%%%
\section{Introduction}

Supersymmetric partition functions have been extensively studied in various contexts and have
led to much insight into the dynamics of these theories.  Typically the
amplitudes depend holomorphically on complex parameters characterizing the theory (through `F-type terms').
A notable example of this is the topological string amplitudes and Nekrasov partition functions which have featured
prominently in the computation of supersymmetric amplitudes on spheres.
Sometimes, however, one encounters more delicate amplitudes which, even though they are characterized
by the same complex parameters, they are {\it not} holomorphic functions of them.
A notable example of this is the $tt^*$ geometry (involving the study of non-abelian Berry-connection for the corresponding
vacua) introduced in the context of ${\cal N}=2$ supersymmetric theories
in 2 dimensions in \cite{cecv} and recently extended to theories with 4 supercharges in $d=3$ and $d=4$ dimensions
\cite{cgv}.  Such amplitudes are in general very complicated functions of the parameters and are not holomorphic.  However,  in very special cases,
for example when the theory is at the conformal point, these amplitudes become simple and factorize into a finite number of
holomorphic/anti-holomorphic blocks.  Another example of supersymmetric amplitudes which are only a function of F-type terms but depend in a complicated way on these parameters is the expectation value of certain loop operators studied by GMN \cite{GMN1,GMN2}.
Both the $tt^*$ amplitudes and the GMN amplitudes depend on a twistor parameter $\zeta$, which is a phase in the physical
region of parameters.

In this paper we uncover a surprising relation between these two types of amplitudes:  We find that in a
particular (somewhat unphysical limit) as $\zeta \rightarrow 0$, the 3d $tt^*$ amplitudes and a suitable 
notion of 5d $tt^*$ amplitudes,  become holomorphic and reduce to open (for $d=3$) and closed
(for $d=5$) A-model topological string amplitudes (and a similar statement for 2d/4d $tt^*$ geometry
and B-model open/closed topological string amplitudes).  These holomorphic functions can in turn be viewed as the building blocks
for the partition functions for these theories on $S^3$ and $S^5$ (or $S^2\times S^1$ and $S^4\times S^1$).   The inverse operation is not simple:
From these holomorphic blocks it is not easy to reconstruct the $tt^*$ amplitudes.  In this sense,
the $tt^*$ amplitudes may be viewed as more fundamental.   Moreover to define the $tt^*$ amplitudes
one does not need to include unphysical additions to the actions to make the theory supersymmetric, unlike
what one does in the context of partition functions on spheres for non-superconformal theories.
Similarly if one considers the other pole of the twistor sphere, given by $\zeta \rightarrow \infty$ we get
the anti-topological string amplitudes.  

Viewing the 5d theory compactified on a circle
as an ${\cal N}=2$ supersymmetric theory in 4 dimensions, we make contact with GMN setup.
In particular we find
an interpretation of GMN line operator amplitudes as giving difference equations for twistorial
closed topological string amplitudes.  This is a twistorial extension of the familiar statement
in the context of topological strings that taking branes around cycles of CY changes the moduli of CY.
In the purely 4d setup, we take the space-time at infinity to be ${\bf R}^2\times {\bf T}^2$ and to define
the amplitudes for this geometry involves the 4d generalization of 2d $tt^*$ to a product of two semi-infinite cigars $C_{1/\epsilon_1}\times C_{1/\epsilon_2}$, where the radii of the cigars at
infinity is related to inverse of the refined topological string coupling constants.\footnote{The amplitudes for this geometry is expected to
be related to that of Taub-Nut by a formal analytic continuation of radii of ${\bf T}^2$.  The ${\cal N}=2$ theory on Taub-NUT was
studied by Neitzke \cite{neitzke} in trying to explain the existence of a canonical line bundle on the moduli space
of the associated hyperK\"ahler metrics (see also \cite{pio}).}

The organization of this paper is as follows:  In section 2 we give a brief review of aspects of $tt^*$ geometry in 2 dimensions.  In section 3 we review the 3d $tt^*$ geometry and its relation to blocks
for the partition function of supersymmetric theories \cite{cgv}.  Moreover we explain, in the context of 3d theory arising from
M5 branes wrapped on Lagrangian CY cycles, how this relation maps this limit of $tt^*$ amplitudes to
open topological string amplitudes.  In section 4 we motivate 5d and 4d versions of $tt^*$ geometry and show
how it reproduces the closed A and B topological string amplitudes in a certain limit.   In section 5 we discuss how the GMN
line operators can lead to difference equations for the twistorial closed topological string amplitudes.  In section 6
we give some examples of how this relation works.  In section 7 we discuss some directions for future work.

\section{Review of aspects of $tt^*$ geometry in 2 dimensions}

In this section we review aspects of $tt^*$ geometry in 2 dimensions, focusing
on aspects relevant for this paper.  For more detail see the original paper \cite{cecv} and a recent
review \cite{cgv}.

Consider a $(2,2)$ supersymmetric theory in 2 dimensions, which admits deformations with mass gap.
This theory will typically have a number of degenerate vacua.
$tt^*$ geometry studies the geometry of this bundle of vacua and the Berry's connection it comes
equipped with, as we change the parameters of the theory.
In addition these equations can be formulated as flatness condition for an improved connection.
Moreover the sections of this flat connection are D-brane amplitudes for this theory.
To be more specific, let $\phi_i$ denote elements of the chiral ring.  Consider deformations of the
theory given by addition of $t^i \phi_i$ to the superpotential $W$.  Denote the action of $\phi_i$
on the vacuum subector by $C_i$.  Furthermore, let $A_i$ denote the Berry's connection for
the vacuum bundle and $D_i$ the corresponding covariant derivatives.   The equations characterizing the connection $A_i$
can be captured by introducing a formal parameter $\zeta$ and defining modified
connections by
$$\nabla_i=D_i+\zeta^{-1} C_i \qquad {\overline \nabla_i}={\overline D_i}+\zeta {\overline C_i}$$
Then the $tt^*$ equations are simply the flatness conditions:
$$[\nabla_i,\nabla_j]=[\nabla_i,{\overline \nabla_j}]=[{\overline \nabla_i},{\overline \nabla_j}]=0.$$
The existence of these improved connections raise the question of what correspond to the flat sections of it.
It turns out that the sections are wave functions associated with D-brane boundary states \cite{hiv}.  These
D-brane boundary conditions are characterized by which $N=1$ supersymmetry we preseve in the $N=2$ theory
which in turn is characterized by a phase $\zeta =e^{i\phi}$.  Let $|D^a(\zeta)\rangle$ be such a state.  Then
$$\Psi^a (t^i,\overline{ t^i},\zeta)=\langle 0|D^a(\zeta)\rangle$$ 
turn out to be the flat sections of the $\nabla_i, {\overline \nabla_i}$ \cite{hiv}.  To define vacuum state $\langle 0|$
one can perform the path-integral on the semi-infinite cigar, which is topologically (or anti-topologically) twisted.

In general the wave functions $\Psi^a$ are complicated functions of the variables and cannot be in general
computed in a closed form.  There are two exceptions:  First, if the 2d theory is at the conformal point, it
turns out that $\Psi^a$ become holomorphic (and anti-holomorphic) sections solved in terms of period integrals.
When the 2d theory is given by a LG theory with a superpotential $W$, this happens when $W$ is quasi-homogeneous,
and one finds:
$$\Psi^a(t^i,\zeta)=\int_{D^a} dx^I\  {\rm  exp}[\zeta^{-1} W(x^I, t^i)]$$
However, generically $W$ is not quasi-homogeneous, and in such cases $\Psi^a$ is a very complicated
function and there is no simple closed formula for it.
There is however an unphysical limit where the expressions dramatically simplify.  For this purpose it is convenient
to introduce the circumference of the cigar $R_c$.  This has the effect of $W\rightarrow R_c\cdot W$.  
To get rid of the complicated mixture of $t,{\overline t}$ in the wave function we consider an {\it unphysical} limit
by taking $\zeta\rightarrow 0$.  This is unphysical because the physical D-brane states require
$\zeta$ to be a phase\footnote{For a conformal theory, if we take $\zeta$ not to be a phase,
an asymmetric conforaml rescaling of chiral versus anti-chiral fields can transform $\zeta$ back to a phase,
which is why in the conformal case we can take $\zeta$ to have arbitrary norm.} and in particular $|\zeta|=1$.  Another way to see this
is that the effect of $\zeta$ can be undone by the transformations
$$W\rightarrow \zeta^{-1} W\quad {\overline W}\rightarrow \zeta {\overline W},$$
which for a unitary theory requires $\zeta^*=\zeta^{-1}$.  Nevertheless the $tt^*$ equations make
sense for arbitrary complex parameter $\zeta$ and we can analytically continue the wave function of the
D-brane to this unphysical value.  In such a case, by taking the limit
$$R_c\rightarrow \epsilon R_c ,\quad \zeta \rightarrow \epsilon,\quad  \epsilon \rightarrow 0$$
%c
one finds that the D-brane wave function become purely holomorphic \cite{hiv}:
$$\Psi^a(t^i,\overline{t^i},\zeta\rightarrow \epsilon, R_c\rightarrow \epsilon R_c)=\Psi^a(t^i,R)=\int_{D^a} dx^I\  {\rm  exp}[{ R_c} W(x^I, t^i)].$$

Despite the simplicity of the final answer in this limit, we would like to emphasize that this is an unphysical limit.
As we will see later in this paper the topological string amplitudes can be recovered from such unphysical analytical
continuations from a physically well defined object.

Recently the partition function of supersymmetric $(2,2)$ theories on $S^2$ has been studied
\cite{B1,D1,H1}.   In order to make the theory on $S^2$ supersymmetric, certain terms needed to be
added to the action, which were unphysical and in particular violate unitarity for the non-conformal case.
Nevertheless the partition function was computable and it was found that the partition functions take
the form
$$Z_{S^2}=\sum_i \chi^a(t^i) {\overline \chi_a}({\overline t^i})$$
 Moreover it was found that 
$$\chi^a(t)=\int_{D^a} exp(W)=\Psi^a(t,\overline t, \zeta)\bigg|_{\zeta\rightarrow 0}$$
In other words in the same unphysical limit where the D-brane wave functions simplify, the unphysical
$S^2$ partition functions are made up of these simplified blocks.  In the conformal case no limit
needs to be taken and the physical partition function on $S^2$ is made up of the physical
blocks which simplify in the conformal limit, as noted before.  This naturally raises the question of whether there
is a more `physical' partition function for non-superconformal theories.  There indeed is such
a partition function and that was the subject of $tt^*$ geometry \cite{cecv} where one considers
infinitely elongated sphereical partition funcation which can be decomposed to two infinite cigars where
the partition function is opposistely twisted on the two cigars.  In this formulation the partition function becomes\footnote{The choice of the phase $\phi$ drops out from the sum in this final expression.}
$$Z_{S^2}^{tt^*}=\sum_a \Psi^a(t^i,{\overline t^i},\zeta =e^{i\phi})\Psi^*_a(t^i,{\overline t^i},\zeta=-e^{i\phi})$$
and is made of a finite number of blocks again, but the blocks are highly complicated functions of $t^i, {\overline t^i}$.
The payoff is that the objects are now made of physical D-brane wave functions where $\zeta=e^{i\phi}$ is a phase (and
the choice of phase disappears in the combination above for the partition funciton on $S^2$).  Moreover this
gives rise to a metric which captures the physical Berry's connection for the vacua of the theory.
  
To summarize, we have seen that the supersymmetric partition functions on $S^2$ in the conformal case can be
made up of physical blocks of D-brane wave functions, however in the non-conformal case the recent definitions
of supersymmetric partition functions even though simple and computable by localization techniques, are not the natural
physical objects to study.  Nevertheless there is an alternative physical object motivated by $tt^*$ geometry which does agree with the supersymmetric partition function in the conformal case but differs from it in the non-conformal case, and is
made of physical D-brane wave functions.  It is natural to ask if the same occurs for higher dimensional
theories.  Indeed, as was pointed out in \cite{cgv} there is a parallel to this in 3d.  Again
one can define a $tt^*$ geometry and brane wave function which in an unphysical limit lead to the building
blocks for supersymmetric partition function, but there also exists a more physical definition of partition functions
which is far more complicated in the non-conformal case.

\section{3d supersymmetric amplitudes and open twistorial topological strings}

The 2d $tt^*$ setup can be extended to theories in 3 dimensions with four supercharges \cite{cgv} (see also
\cite{bdp}).
The basic equations one obtains involves generalizations of Bogomolnyi equation which
can be viewed as reductions of the equations for hyper-holomorphic connections.  The way this was derived
was by viewing the compactification of the three dimensional theory on a circle of radius $R$,  leading
to a 2d theory with infinitely many fields.
In particular for theories in 3 dimensions the Berry connection can again be formulated as flatness
condition by introducting the formal variable $\zeta$.  Moreover when $\zeta$ is a phase, the
flat sections correspond to wave functions for D-brane boundary conditions.  The corresponding
wave functions correspond to path integrals on infinitely elongated cigar of radius $R_c$ times
a circle of radius $R$.  Again as in the 2d case the wave function is a very complicated function
of parameters, but the answer simplifies when one considers the unphysical limit
$$\zeta\rightarrow \epsilon,\ R_c\rightarrow \epsilon R_c, \ \epsilon \rightarrow 0.$$

  We now connect this limit of $tt^*$ geometry
with 3d supsersymmetric amplitudes and to open topological strings. It is known that supersymmetric partition functions
for ${\cal N}=2$ theories in 3 dimensions on $S^3$ or $S^2\times S^1$  is made of holomorphic blocks \cite{bdp} which can be viewed as partition functions
on a solid torus.  This is also consistent with its relation with open topological string amplitudes as we will now explain.

We first recall the embedding of refined closed topological string realizing Nekrasaov's partition function \cite{nek} in M-theory \cite{gopv1,gopv2,dvv,hoiv,acdkv} and then how the open topological
string gets defined in this setup.
Refined A-model closed topological strings is defined in terms of M-theory by considering the background
$$M={\bf C}^2\ltimes S^1\times X$$
 In the
context of string dualities it is natural for ${\bf C}^2$ to have the
 Taub-NUT metric.  However, for what we need in this paper, only the $U(1)\times U(1)$ rotation symmetries
 are relevant and as such we denote it by ${\bf C}^2$. 
 $X$ is the Calabi-Yau and as we go around $S^1$ we rotate the two planes of ${\bf C}^2$ by angles $(\epsilon_1,\epsilon_2)$.
Focusing on the geometry far away from the origin of each ${\bf C}$, or equivalently focusing
on the ${\bf C^*}^{2}$ geometry, there are three
natural circles in our problem:  One is the 5d circle, and in addition there are two circles each associated to the
circle rotation of a plane in ${\bf C}^2$.  Consider the two tori, consisting of the 5d circle and one of the circles associated
to rotation of the planes of ${\bf C^*}^2$.
Let us denote the complex structure of the two tori as $(\rho,\tau)$. The fibration data above implies that
$ {\rm Re}(\rho )=\epsilon_1/2\pi, {\rm Re}(\tau)=\epsilon_2/2\pi$.

The open topological string gets embedded in this setup by considering M5 branes wrapping (special) Lagrangian
cycles $L$ of $X$, wrapping the $S^1$ and a plane in ${\bf C}^2$.  There are two natural choices
of the planes of ${\bf C}^2$ one can pick, compatible with the circle actions.  In the context
of the unrefined topological string, it is natural to take ${\bf C}^2=TN$.  In this case
 the M5 brane geometry as a subspace of ${\bf C}^2$ is an infinite cigar where the radius of cigar $R_c$ is
identified with the asymptotic radius of the circle of the $TN$.  The boundary of M5 brane is thus the same $T^2$
as above with complex moduli $\tau$.  In other words the geometry of the M5 brane is
$$M5:  {\bf R}^+\ltimes T^2\times L$$
where ${\bf R}^+$ is half-line where at its origin the TN circle shrinks.  Viewed from the perspective  
of 3d theory living on the M5 brane, we have an ${\cal N}=2$ supersymmetric theory
whose geometry is $C\times S^1$ where $C$ is a cigar geometry.  But this is precisely
the $tt^*$ geometry we have discussed above, where we identify 
$$\tau=i {R/R_c},$$
where $R$ is the circumference of $S^1$.  This suggests that open topological string and building blocks of suersymmetric partition functions
should be related to $tt^*$ blocks.  Indeed it was shown in \cite{cgv} that at least in simple examples this expectation is realized.
Here we conjecture that this is generally true.  Namely in the same unphysical limit
we outlined above where $\zeta\rightarrow 0$ one would find that blocks of $tt^*$ brane amplitudes
reduce to open topological strings/supersymmetric partition function blocks.  The two geometries
differ in that in one case $\tau=iR/R_c$ and in the other case ${\rm Re}(\tau)=\epsilon_2/2\pi$.   The identification
found in \cite{cgv} was that if one views the topological string coupling constant as $\lambda =i\epsilon_2= 2\pi i\tau$, instead of $i\epsilon_2= 2\pi i{\rm Re}(\tau)$ (which is natural
since the topological string amplitudes depends holomorphically on $\epsilon_2$), the two amplitudes match in this limit.\footnote{Similar observations
were made in \cite{bdp} in connecting the squashing parameter $b^2$ of $S^3$ to modulus $\tau$
of the $T^2$ on ${1\over 2}S^3$.}
 In other words, this
leads to the identification $\epsilon_2=2\pi \tau=2\pi i R/R_c$.  Surprising as this statement sounds, this phenomenon has already been observed and beautifully explained in \cite{nw}, where the role
of rotation parameter in the $\Omega$-deformed theory is exchanged, by a change of variables in the path-inegral to the circumference
of the cigar for the undeformed theory.  In this paper we will be assuming that the same ideas explored there apply to the case at hand.  
For general $\zeta$ (i.e. not necessarily $\zeta \rightarrow 0$), the above $tt^*$ amplitudes is our definition of {\it twistorial open topological string amplitudes}.

This setup naturally raises the question concerning the closed topological string setup.  
In particular if the Calabi-Yau
$X$ is non-trivial (by which we mean it has non-trivial 4-cycles) the 3d theory living on the unwrapped directions of  M5 brane
will be coupled to some bulk theory in 5 dimension, resulting from the compactification of M-theory on Calabi-Yau $X$.
The corresponding $tt^*$ geometry for this coupled system can still be formulated, as discussed in \cite{cgv}.
It is natural to ask what would the limit of $\zeta\rightarrow 0$ corresponds to in this context?  Clearly the
open topological string already includes higher genus amplitudes which probe the non-trivial geometry of the CY.
So it is natural to continue expecting that regardless of whether the CY is simple or not, this limit leads
to twistorial extension of topological string amplitdues including {\it mixed} contributions of closed and open topological strings. 

It is  natural to ask if we can identify the analog of purely closed twistorial topological string.
From the discussion above
in the context of open string, we are naturally led to consider the geometry ${\bf R^2}\times {\bf T^2}$ were the radii
of the ${\bf T}^2$ are (up to a factor of the 5d circumference $R$) given by $(1/\epsilon_1,1/\epsilon_2)$.  We will take
this up in the next section.

Before turning to the bulk 5d version of $tt^*$ geometry there is a related system to the one discussed
which naturally fits with topological B-model.  Indeed the above 3d-5d coupled system is the one higher dimensional version of a similar 2d-4d system studied in \cite{GMN24}, where the corresponding $tt^*$ geometry was studied. This setup
has a natural embedding in string theory involving type IIB strings on Calabi-Yau, with B-branes corresponding
to D3 branes wrapping a 2-cycle in Calabi-Yau and filling a 2d subspace of 4-dimensional space.
  Again,
in that context given the relation of topological A-model with topological B-model (for example
in the context of geometric engineering) one would expect that the same $\zeta \rightarrow 0$ should
lead to open topological B-model amplitudes.

\section{5d $tt^*$, supersymmetric amplitudes and closed twistorial topological strings}

It is natural to try extend the $tt^*$ geometry to theories with $8$ supercharges, and in particular
to ${\cal N}=1$ supersymmetric theory in 5 dimensions and ${\cal N}=2$ supersymmetric theory 
in 4 dimensions.  As we saw in the last section the connection
with open topological string strongly suggests that this exists.   In fact the idea is to use exactly the same setup:
In the context of 2d $tt^*$ we take the Hilbert space to be on a circle and we study the amplitudes of the D-brane
by considering the path-integral on an infinite cigar geometry, where the circle is the boundary of cigar.
Similarly in the 3d $tt^*$ we have a $T^2$ geometry and one circle gets filled in as in the cigar geometry
(times an additional circle).  

The embedding of the open topological string in the closed topological string
discussed in the last section suggests that in the same context we have to consider a boundary
condition, i.e. a D-brane for the 5d theory.  So to define the closed string analog
of $tt^*$ we repeat what we did for the 3d and 2d case:  We define a D-brane, i.e. a boundary condition on the product
of two (R-twisted) cigars, $C_1\times C_2$,
which preserves half of the supersymmetry.  Of course as in the 2d case we need an infinitely long neck.  Moreover viewed from the viewpoint
of the 4d ${\cal N}=2$ theory, we have to choose a phase $\zeta$
which determines which combinations of supersymmetry we wish to preserve.  Note that this exists since
supersymmetric boundary conditions on the Coulomb branch with boundary $R^3$ exist and far away from the origin
the space becomes effectively the flat geometry.\footnote{In particular the boundary
conditions on the bosonic fields as we approach the infinity takes the form 
$${\bf E}_i+i{\bf B}_i=\zeta^{-1}\nabla \phi_i, \qquad 
\phi_i\rightarrow a_i$$
where $\phi_i$ are scalars in the vector multiplets and $\zeta$ is a phase.}
  Doing the path-integral
on this geometry we get a partition function which we call the `twistorial topological string'
$Z^{tt^*}(X^i,{\overline X^i},\zeta, \epsilon_i)$ where $X^i$ are the (complexified) K\"ahler moduli of the Calabi-Yau $X$
and $\epsilon_i$ are propotional to inverse of the circumferences of the cigars.
One can also consider turning on Wilson lines on the circles which makes the partition function
include a twist parameter (as in 3d $tt^*$) which in this case makes the moduli space hyperK\"ahler. For simplicity
of exposition we set these parameters to zero for the rest of this paper.
We expect that the topological string partition function $Z(X^i,\epsilon_i)$  is recovered in the limit
we found in the context of open strings, namely
$$Z^{top}(X^i,\epsilon_1,\epsilon_2)=Z^{tt^*}(X^i,{\overline X^i},\zeta \rightarrow \epsilon, \epsilon_i \rightarrow \epsilon_i/ \epsilon )\bigg|_{\epsilon \rightarrow 0}$$

That such an object should exist and this relation between twistorial version and the usual topological
string should work is supported by extending the relation we have for the open topological strings to closed topological strings.
Nevertheless the question remains as to how one can compute these amplitudes.  In the next section we show
how one can use the line operator amplitudes computed by GMN to obtain difference equations for the twistorial topological string amplitudes.

One can also ask what the purely 4d version of this $tt^*$ is.  The idea would be simply to consider the geometry
$C_1\times C_2$ putting a boundary condition
preserving the supersymmetry dictated by $\zeta$.   The amplitude should preserve the supersymmetry in the limit the boundary is taken towards infinity, as in the 2d $tt^*$.  In this context the same limit
should lead to B-model topological string amplitudes with $\epsilon_i=i /R_{C_i}$.
The compactification of ${\cal N}=2$ theories on $TN$ was studied in \cite{neitzke} in the context
of showing the existence of a hyperholomorphic line bundle on the moduli space of circle compactification
of these theories (see also \cite{pio}).  Moreover it was suggested that there should be a natural section for this bundle\footnote{It was
pointed out to me by Neitzke that one motivation for this question was in fact to relate this section to the partition function of topological strings.}.  It seems natural to identify this line bundle with the vacuum bundle (at least when $\epsilon_1+\epsilon_2=0$).  Since the vacuum is one dimensional (for a fixed Coulomb branch point) this is a line bundle.  Moroever doing the path-integral picks a state in the Hilbert space, whose overlap with the boundary state (depending on $\zeta$) can be viewed as a section for this bundle which leads to the closed twistorial
topological string partition function.

\section{Recursion relations for twistorial topological strings}
  In this section we show how the GMN line operator amplitudes can be used to deduce difference equations
for closed twistorial topological strings.  The basic idea is to recall an analogous fact about the relation between
open topological string amplitudes as leading to difference equations for closed topological strings and to show
that the twistorial version of these equations give the desired relation between GMN line operators and difference
equations for closed topological strings.  Our discussion in this section will be somewhat heuristic
and in the context of the unrefined topological string where $\lambda =\epsilon_1=-\epsilon_2$.  The discussion becomes more precise
when in the next section we extend it to the refined topological string.

Consider B-model topological string.  Consider in addition a B-brane wrapping a 2-cycle.  In a local
model where the CY is represented by a curve, the B-brane is represented by a point on the curve.
The following phenomenon is well known in the context of topological strings \cite{ADKMV,acdkv}.
If one considers taking the B-brane around a cycle and back to itself, spanning a 3-cycle $\gamma$ in the process,
the partition function of topological string does not come back to itself\footnote{More precisely one considers
adding a brane/anti-brane system and take the brane around the cycle before annihilating it with
the anti-brane as in Verlinde's work \cite{ver}.}.  Instead it gets multiplied by
$$U_\gamma Z^{top}=exp({2\pi i a_\gamma \over \lambda} ) \cdot Z^{top}$$
where $a_{\gamma}$ are viewed as operators with commutation relations.  Namely the
partition function of topological string $Z^{top}$ should be viewed as  a wave function in the Hilbert space \cite{with}
(as an expression of holomorphic anomaly \cite{bcov}), with the commutation relations for $a_{\gamma}$ given by
$$[a_{\gamma},a_{\gamma '}]= \langle \gamma, \gamma'\rangle {\lambda^2\over 2\pi i}$$
In particular taking a symplectic basis we can use $a_i, a^D_i$ as canonically paired basis elements and view the topological
string as a function of the $a_i$ variables: $Z^{top}(a_i,\lambda)$.
Then 
$$U_{a_i^D}=exp(\lambda d/da_i)$$
Acting this on $Z^{top}$ we get
$$U_{a_i^D }Z^{top}(a_i,\lambda)=Z^{top}(a_i+\lambda,\lambda)=\chi_{i^D}(a_i,\lambda) Z^{top}(a_i,\lambda)$$
where
$$\chi_{i^D}={Z^{top}(a_i+\lambda,\lambda)\over Z^{top}(a_i,\lambda)}$$
In other words we have a difference equation:
$$Z^{top}(a_i+\lambda,\lambda)=\chi_{i^D} \cdot Z^{top}(a_i,\lambda)$$
These difference equations fix $Z^{top}$ up to non-perturbative terms
$$Z^{top}\rightarrow Z^{top}\sum_n c_n \ {\rm  exp}(2\pi in_i a^i/\lambda)$$

The question we wish to address is what is the twistorial version of this equation?  We now argue
that the same type of equation works with the identification
$$\chi_\gamma\leftrightarrow X_\gamma$$
where $X_\gamma$ are the vev of line operators studied in \cite{GMN1}.  This in fact simply follows
from the fact that in the 2d-4d system studied in \cite{GMN24} if one takes the 2d surface
operators around a cycle $\gamma$ we pick up a multiplier $X_\gamma$, modulo the choice
of the refinement parameter discussed in the previous section.   Let us ignore this subtlety for
the rest of this section and return to it in the following section.  Here we discuss whether this
identification is at least on the right track.

 Since the corresponding
objects reduce to the B-branes and their monodromies as we take the $\zeta\rightarrow 0$ limit, we obtain
the above correspondence.
Similarly these difference equations fix the twistorial topological string amplitudes up to non-perturbative terms.
More specifically we view the 5d theory compactified on a circle of radius $R$ as a 4d system, taking
into account the KK modes.   For this system we can study the GMN line operators $X_\gamma$ which 
will be a function of Coulomb branch parameters $a^i$, the twistor parameter $\zeta$ and the radius $R_c$ of
the circle we compactify as part of the 4d theory and so the vev of the line operators are functions of the form
$X_\gamma (a^i,{\overline a^i},\zeta , R,R_c)$.  We have
for large $R_c$ 
$$X_\gamma \sim exp\bigg[-{R_c Z_\gamma(a^i)\over 2\pi   R \zeta}-{\zeta R_c\overline {Z_\gamma(a^i)}\over 2\pi  R}\bigg]
\sim exp\bigg[{Z_\gamma(a^i)\over \lambda \zeta}+ {\zeta \overline {Z_\gamma(a^i)}\over {\lambda} }\bigg]$$
where $Z_\gamma$ denotes the central charge associated to the charge sector $\gamma$  and we take $\lambda ={-2\pi  R\over R_c}$; the extra factor of $1/R$ compared to GMN is due to the normalization of the energy coming from the 5d theory
compactified on a circle of radius $R$.  Then we get the difference equations for the twistorial topological strings as
$$Z^{tt^*}(a^i+\lambda \zeta ,\overline{a^i}+{ \lambda}\zeta^{-1},\zeta ,R,R_c)=X_{i^D}(a^i,{\overline a^i},\zeta , R,R_c)
\cdot  Z^{tt^*}(a^i,\overline {a^i},\zeta ,R,R_c)$$
where $i^D$ denotes the magnetic charge dual to $i$.
For the large $R_c$ regime using the asymptotic form of $X_\gamma$ we can solve for $Z^{tt^*}$:
$$Z^{tt^*}\sim exp\bigg({ F(a^i)\over \lambda^2 \zeta^2}+{{\overline {F(a^i)}}\zeta^2\over { \lambda}^2} \bigg)$$
where $F(a^i)$ is the prepotential.  This is consistent with the fact that $a^{iD}=\partial F(a_i)/\partial a_i$
and that shift of $a^i$ is proportional to $\lambda$ which is small in this limit.
The usual topological string is recovered in the unphysical limit of $\zeta \rightarrow \epsilon, \lambda \rightarrow \lambda /\epsilon,
\epsilon \rightarrow 0$, leading to the large $R_c$, small $\lambda$ limit
$$Z^{tt^*}\rightarrow Z^{top}\sim exp\big(F(a^i)/\lambda^2\big)$$
which indeed agrees with the expected form of the topological string partition function for small $\lambda$.
This shows that at least heuristically things work in a nice way.  However, to be more precise we
have to take note of the refinement of the topological strings, which we now turn to.

\subsection{Twistorial Extensions for the Refined Topological Strings}

So far we have talked about ordinary topological string amplitudes.    However to make this
dictionary precise we need to take into account the refinement of topological string.

In the refined case there are two distinct M5 branes we can use, one wrapping the $z_1$ plane and the other
the $z_2$ plane.    Taking such branes around cycles of Calabi-Yau, leads to operators acting on the
topological string partition function by
$$U^i_\gamma Z^{top}=exp({2\pi i a_\gamma \over \epsilon_i} ) \cdot Z^{top}$$
where $a_{\gamma}$ are viewed as operators with commutation relations.  Namely
$$[a_{\gamma},a_{\gamma '}]= \langle \gamma, \gamma'\rangle {\epsilon_1\epsilon_2\over 2\pi i}$$
Let us fix it to be the $z_2$ plane, leading to $U^2_\gamma$ operators.  In this case we find that $U^2_\gamma, U^2_{\gamma'}$
do not commute in general, and lead to
$$U^2_\gamma U^2_{\gamma'}= \alpha U^2_{\gamma'} U^2_\gamma $$
where
$$\alpha =exp\big[2\pi i \langle \gamma,\gamma'\rangle {\epsilon_1\over \epsilon_2}\big].$$
From what we have said before, we would like to identify $U^2_{\gamma}$ with the GMN line operators when
we extend the theory to the twistorial setup with some choice of the parameter $\epsilon_1$.  There
are infinitely many choices of the parameter $\epsilon_i$ which makes the $X_\gamma$ operators commutative:
$\epsilon_1=n\epsilon_2$, for some integer $n$.    The one corresponding to GMN should be the $\epsilon_1=0$
limit, i.e. the Nekrasov-Shatashvili limit \cite{NS}, because in that limit one circle of the two cigars becomes of infinite size and we get the geometry $S^1\times R^3$ which is
the geometry studied originally by GMN.  Thus the original GMN setup should lead to NS limit of topological string
\cite{toapp}.  In this limit the difference equation will become the differential equation as was studied from this
point of view in \cite{acdkv}.
    For this to make sense the GMN line operators should admit a non-commutative refinement.  Indeed such an extension of GMN setup was studied in \cite{GMN2}, where as one goes
around the $S^1$ of the 4d theory one also rotates around an axis in ${\bf R}^3$ perpendicular
to where the brane is wrapped, by an angle $\theta$ and the GMN line
operators become non-commutative:
$$X_\gamma X_{\gamma'}=\alpha_{\gamma,\gamma'}X_{\gamma'}X_\gamma$$
where
$$\alpha_{\gamma,\gamma'}=exp\big[i\langle \gamma,\gamma'\rangle \theta\big]$$
Since in our 5d setup the brane is in the $z_2$ plane this rotation will be in the $z_1$ plane.
To complete this relation we need to show that in the $\zeta\rightarrow 0$ we get the identification
$$\theta =2 \pi {\epsilon_1 \over \epsilon_2} \qquad mod\ 2\pi.$$
There are three circles involved: the two circles of ${\bf C}^2$ which we denote by $1,2$ and
the 5d circle, which we denote by $3$.  The fibration structure of these circles define
a $T^3$.  For each $T^2\subset T^3$ we can use the complex ratio of the complefixied lengths
to denote the structure of the fibration.  Let us denote the corresponding complex vectors $z_i$
where each $T^2$ will have a complex structure $\tau_{ij}=z_i/z_j$.
From the definition of the refined topological string (and again extending real part of $\tau$ to the full $\tau$) we have
$$\tau_{31}={z_3\over z_1}={\epsilon_1\over 2\pi} \qquad \tau_{32}={z_3\over z_2}={\epsilon_2 \over 2\pi}$$
On the other hand from the definition of the rotation of the GMN, since we identify their circle with the
2-circle and since as we go around it we rotate the $z_2$ plane we have
$${\theta \over 2\pi}={z_2\over z_1}={\tau_{31}\over \tau_{32}}={\epsilon_1\over \epsilon_2}$$
which establishes what we wished to show.  This provides further evidence that the refined GMN setup leads to difference equations for twistorial extension of the refined topological string partition function (which should be viewed as a section of a bundle).

\section{An Illustrative Example}
In this section we show how the topological string partition function can be recovered from this twistorial setup in an example.
Conifold (in the B-model) gives rise to a  4d theory with ${\cal N}=2$ supersymmetric theory with $U(1)$ gauge group
with a matter hypermultiplet.  Let the Coulomb branch parameter be denoted by $\mu$.  Let $X_{e,m}$ denote the
electric/magnetic
Wilson loop vevs. Then
from \cite{GMN1} we have
$$X_m= X_m^{sf} exp\bigg[ {i\over 4\pi}\int_{l_+} {d\zeta'\over \zeta'} {\zeta'+\zeta \over \zeta'-\zeta}{\rm log}[1-X_e(\zeta')]$$
$$
{-i\over 4\pi}\int_{l_-} {d\zeta'\over \zeta'} {\zeta'+\zeta \over \zeta'-\zeta}{\rm log}[1-X_e(\zeta')^{-1}]\bigg]$$
for suitable choices of contours $l_+,l_-$ and where 
$$X_e({\zeta '})=exp\bigg[{\mu \over \lambda \zeta'} +{{\overline \mu}\zeta'\over {\lambda}}\bigg]$$
and 
$$X_m^{sf}=exp\bigg[{\mu {\rm log }\mu - \mu\over \lambda \zeta}+ \zeta{{\overline \mu} {\rm log}{\overline \mu}-{\overline \mu}\over { \lambda}}\bigg]$$
To recover the usual topological string we need to take the limit $\zeta \rightarrow \epsilon, \lambda \rightarrow {\lambda \over \epsilon},\epsilon \rightarrow 0$.  As will be discussed in \cite{toapp} one finds in this limit
$$X_m\rightarrow \Gamma ({\mu\over \lambda})$$
in agreement with the fact that in the NS limit ${\rm log}Z^{top}= F/\epsilon_1$ and $\epsilon_2 \partial F/\partial \mu ={\rm log} \ \Gamma(\mu/\epsilon_2)$,
where $\lambda =\epsilon_2$.
 
 Note that the twistorial version above, using the large N duality of matrix
model with topological strings \cite{dv}, leads to a twistorial version of the matrix model.  Namely we view the
matrix model of \cite{dv} as the superpotential of a 2d ${\cal N}=2$ supersymmetric system and define the twistorial
topological string by its D-brane wave functions.
This is currently being studied \cite{toapp}.
It is also interesting to consider the amplitudes away from the NS limit, which is being studied in \cite{cecvapp}.
Both the twistorial and the limiting case can be easily extended to the A-model conifold amplitudes
by considering a sum where $\mu \rightarrow \mu + in $ for $n\in {\bf Z}$ corresponding to the KK modes.
The standard topological string limit gives rise to the usual quantum tri-log and the twistorial version is being studied in \cite{cecvapp}.

\section{Future Directions}
In this paper we have introduced the notion of a twistorial topological string which interpolates between topological
string on the north pole, to anti-topological string on the south pole.  Moreover the more `physical' object corresponds
to the equator of the twistor sphere.    The structure of the amplitude is very smilar to the localization ideas pioneered
by Pestun \cite{Pestun} in the context of supersymmetric partition functions on spheres.  It would be interesting to see if one can formulate this twistorial notion of topological string
(or the twistorial Nekrasov partition functions) in that setup.

It is clear that the twistorial topological string is far more complicated than the usual topological strings, even
though both are characterized by BPS particles. In this paper we have defined them by a path-integral, and
proposed a recursion relation which fixes them up to non-perturbative terms.  It would be desirable to derive
this recursion relation more directly and also fix the non-perturbative terms.  In particular it would be interesting
to connect this to the objects studied in \cite{neitzke} which satisfy a similar difference equation.

In the context of black holes, it was proposed in \cite{osv} that the partition function of the BPS black
holes is the absolute square of the topological string partition function.  Given that the twistorial
topological string is a mixture between topological and anti-topological string and that it
caputres the BPS jumps through the twistor parameter, it would be interesting to see if this leads
to a modification of the OSV conjecture, in terms of the twistorial topological string.

Finally it was proposed in \cite{witten} that one can define path-integral cycles, even for theories
without a positive-definite action, by complexifying the fields and viewing the action as a superpotential of the
field variables and use $ \int_D e^W$ for suitable cycles $D$ (leading to convergent integrals) as definitions
of the path-integral.  This would in particular agree with the usual definition of quantum mechanics when
$-W$ is positive definite.  However, as already noted in this paper, there is another well motivated defintion
of such a wave function which, even though it is far more complicated and includes an additional parameter,
is more natural from the viewpoint of $tt^*$ geometry (roughly speaking we have two independent notions of $\hbar$, one for $W$ and one for ${\overline W}$).  
It would be interesting to see what deformation of quantum mechanics this leads to.

\section*{Acknowledgments}

I have greatly benefitted from discussions with M. Aganagic, S. Cecotti and A. Neitzke.  In addition
I would like to thank C. Cordova, T. Dumitrescu, D. Gaiotto, B. Haghighat, J. Heckman, V. Pestun, T. Rudelius and E. Witten for valuable discussions.
This research was supported in part by NSF grant PHY-1067976.

\end{document}